\documentstyle[preprint,aps,floats,psfig]{revtex}
\catcode`@=11
\def\references{%
\ifpreprintsty
\bigskip\bigskip
\hbox to\hsize{\hss\large \refname\hss}%
\else
\vskip24pt
\hrule width\hsize\relax
\vskip 1.6cm
\fi
\list{\@biblabel{\arabic{enumiv}}}%
{\labelwidth\WidestRefLabelThusFar  \labelsep4pt %
\leftmargin\labelwidth %
\advance\leftmargin\labelsep %
\ifdim\baselinestretch pt>1 pt %
\parsep  4pt\relax %
\else %
\parsep  0pt\relax %
\fi
\itemsep\parsep %
\usecounter{enumiv}%
\let\p@enumiv\@empty
\def\theenumiv{\arabic{enumiv}}%
}%
\let\newblock\relax %
\sloppy\clubpenalty4000\widowpenalty4000
\sfcode`\.=1000\relax
\ifpreprintsty\else\small\fi
}
\catcode`@=12
\begin{document}
%my defs:
\def\mh{m_h^{}}
\def\vev#1{{\langle#1\rangle}}
\def\gev{{\rm GeV}}
\def\tev{{\rm TeV}}
\def\fbi{\rm fb^{-1}}
\def\lsim{\mathrel{\raise.3ex\hbox{$<$\kern-.75em\lower1ex\hbox{$\sim$}}}}
\def\gsim{\mathrel{\raise.3ex\hbox{$>$\kern-.75em\lower1ex\hbox{$\sim$}}}}
\newcommand{\hmu}{{\hat\mu}}
\newcommand{\hnu}{{\hat\nu}}
\newcommand{\hrho}{{\hat\rho}}
\newcommand{\hh}{{\hat{h}}}
\newcommand{\hg}{{\hat{g}}}
\newcommand{\hk}{{\hat\kappa}}
\newcommand{\tA}{{\widetilde{A}}}
\newcommand{\tP}{{\widetilde{P}}}
\newcommand{\tF}{{\widetilde{F}}}
\newcommand{\th}{{\widetilde{h}}}
\newcommand{\tp}{{\widetilde\phi}}
\newcommand{\tchi}{{\widetilde\chi}}
\newcommand{\te}{{\widetilde\eta}}
\newcommand{\vn}{{\vec{n}}}
\newcommand{\vm}{{\vec{m}}}
\renewcommand{\theenumi}{\roman{enumi}}
\renewcommand{\labelenumi}{(\theenumi)}
%%%%%%%%%%%%%%%%%%%%%%%%%%%%%%%%
%%%%%%%%  Slash character...

\newcommand{ \slashchar }[1]{\setbox0=\hbox{$#1$}   % set a box for #1
   \dimen0=\wd0                                     % and get its size
   \setbox1=\hbox{/} \dimen1=\wd1                   % get size of /
   \ifdim\dimen0>\dimen1                            % #1 is bigger
      \rlap{\hbox to \dimen0{\hfil/\hfil}}          % so center / in box
      #1                                            % and print #1
   \else                                            % / is bigger
      \rlap{\hbox to \dimen1{\hfil$#1$\hfil}}       % so center #1
      /                                             % and print /
   \fi}                                             %

\tighten
\preprint{ \vbox{
\hbox{MADPH--01-1224}
\hbox{AMES-HET-01-04}
\hbox{hep-ph/0104095}}}
\draft
\title{Earth Regeneration of Solar Neutrinos at SNO \\ and Super-Kamiokande}
\author{V. Barger$^1$, D. Marfatia$^1$, K. Whisnant$^2$ and B. P. Wood$^1$}
\vskip 0.3in
\address{$^1$Department of Physics, University of Wisconsin--Madison, WI 53706}
\vskip 0.15in
\address{$^2$Department of Physics and Astronomy, Iowa State University, Ames, IA 50011}
\vskip 0.15in
\vskip 0.1in

\maketitle

\begin{abstract}
{\rm We analyze the 1258-day Super-Kamiokande day and night solar neutrino 
energy spectra with various $\chi^2$ definitions.  The best-fit
lies in the LMA region at 
$(\Delta m^2,{\rm{tan}^2}\theta)=(5.01\times 10^{-5}\,{\rm{eV}^2},0.60)$, independently of
whether systematic errors are included in the $\chi^2$-definition. 
We compare the exclusion and allowed regions from the different 
definitions and choose the most suitable definition to predict the regions
 from SNO at the end of three years of data accumulation. We first work under the
assumption that Super-Kamiokande sees a flux-suppressed flat energy spectrum.
Then, we consider the possibility of each one of the three MSW regions being the solution
to the solar neutrino problem. We find that the exclusion and allowed regions for the 
flat spectrum hypothesis and the LMA and LOW solutions are alike.
In three years, we expect SNO to find very similar 
regions to that obtained by Super-Kamiokande. 
We evaluate whether the zenith angle distribution at SNO with optimum binning 
will add anything to the analysis of the day and night spectra; for comparison, 
we show the results of our analysis of
the 1258-day zenith angle distribution from Super-Kamiokande, for which 
the best-fit parameters are 
$(\Delta m^2,{\rm{tan}^2}\theta)=(5.01\times 10^{-5}\,{\rm{eV}^2},0.56)$. }
\end{abstract}

\section{Introduction}

Neutrinos oscillate. Atmospheric neutrino experiments~\cite{k1,sk1,imb,soudan,macro} provide
compelling evidence for this. The solar neutrino problem~\cite{bahcall} 
has been in existence for thirty years, long before the first indications of an atmospheric 
anomaly. Various solar neutrino experiments~\cite{homestake,sage,gallex,gno,sk,s} detect a 
flux-deficit of  $1/2$ to $1/3$ of the Standard Solar Model (SSM) prediction \cite{SSM}. 
The deficit can be explained
 by invoking the neutrino oscillation hypothesis. 
Despite this, the solar neutrino problem is unsolved. Super-Kamiokande (SK)~\cite{sk,s} 
has not found evidence for any of 
the three litmus tests for neutrino oscillations: energy spectrum distortion, zenith angle 
dependence of the flux (arising from the earth regeneration effect ~\cite{bppw,mikh,baltz}) 
or seasonal variations of the flux. Moreover, there are three distinct robust
solutions all of which have comparable significance levels, LMA, SMA and 
LOW~\cite{analysis,sol}. A VAC solution at 
\mbox{$\Delta m_{21}^2\approx 10^{-10}$ $\rm{eV}^2$} 
is fragile in comparison to the three other solutions and its existence depends sensitively upon
how much emphasis is placed on the SK data~\cite{analysis}. We do not consider it 
further and focus on the MSW solutions.
 Presently, data from SK favors the LMA 
solution~\cite{s}. 
The KamLAND reactor neutrino experiment~\cite{kamland} will establish once and for all 
whether or not the LMA solution is correct, independent of the solar neutrino flux. 
If it is, within three years we will know $\Delta m_{21}^2$ 
 and $\rm{sin}^2 2\,\theta_{12}$ to an accuracy of $\pm 10$\% and $\pm 0.1$, respectively~\cite{kam,comp}.
The SNO experiment~\cite{sno}, too, is expected to make significant inroads towards the resolution of the
solar neutrino puzzle~\cite{smirnov,snozenith,anal}, especially through neutral current 
measurements.

%\section{The Sudbury Neutrino Observatory}
SNO is a Cerenkov detector with 1000 tons of heavy water as its detection medium. Its central
 objective is to test whether electron neutrinos produced in the Sun 
oscillate into active or sterile neutrinos. This can be accomplished by the simultaneous 
measurement of the rates of 
the charged current (CC) and neutral current (NC) reactions
\begin{eqnarray}
\nu_e + d &\rightarrow& p+p+e^-\ \ \ \ \ \ (\rm{CC}) \label{cc}\\
\nu_x + d &\rightarrow& p+n+\nu_x'\ \ \ \ \ \ (\rm{NC})
\end{eqnarray}    
where $\nu_x$ denotes any of the active flavors. The NC reaction measures the total flux of
active neutrinos which is the same as the $^8$B flux produced in the Sun 
if there are only active-active
 neutrino oscillations. Thus, a cross section-normalized ratio NC/CC $\sim 2.5$ 
indicates the oscillation 
\mbox{$\nu_e \rightarrow \nu_{\mu}/ \nu_{\tau}$.} 
If oscillations into sterile neutrinos occur, both the CC and NC rates will be 
suppressed giving NC/CC of unity thereby signaling the existence of sterile neutrinos. 
Since the energy threshold is expected to be about 5 MeV for the CC reaction and 2.2 MeV for the 
NC reaction, only $^8$B and $hep$ neutrinos will contribute to the SNO event rates. 

Because the NC reaction
is unique to SNO, a number of studies have been devoted to its 
exploitation. In this work we undertake an analysis of what the CC rate measurements 
at SNO by themselves can and cannot tell us.
They can provide valuable information for active-active 
oscillations but are not as sensitive to active-sterile oscillations. 
The recoil energy spectrum of CC events and their
zenith angle distribution can in principle  eliminate two of the three globally allowed regions
in oscillation parameter space, 
and also measure the oscillation parameters. 

The stringent CHOOZ limit~\cite{chooz} (see also The Palo Verde Experiment~\cite{paloverde}) 
of \mbox{$\rm{sin}^2 2\,\theta_{13} < 0.1$} 
%for $\Delta m_{21}^2 \gsim 10^{-3}$ $\rm{eV}^2$ 
(at the 95\% confidence level), approximately decouples solar neutrino oscillations
from atmospheric neutrino oscillations.
For small values of $\theta_{13}$, provided $\Delta m_{21}^2 \ll \Delta m_{32}^2$, 
the three-flavor survival probability $P_3$ is related to the  
two-flavor survival probability $P_2$ by (see the first paper of Ref.~\cite{sol}),
\begin{equation}
P_3 \simeq {\rm{cos}}\, 2\,\theta_{13}\ P_2 \ \ \ \Rightarrow \ \ \  0.95\, P_2 \lsim P_3 \leq P_2\,,
\end{equation}
where the inequality arises from the CHOOZ limit on  $\theta_{13}$. 
Thus, even when the limit is saturated, the
two-neutrino analysis represents a very good approximation to the three-neutrino analysis. 
Our analysis is performed in the two active neutrino framework. 

In \mbox{Section II} we briefly describe how the electron recoil energy
spectra expected in the daytime and nighttime are calculated. 
We will collectively call both spectra the ``D\&N spectra''.
In Section III we analyze the D\&N  data from SK (1258 effective days) 
with different $\chi^2$ definitions and find the optimum definition for
the analyses of the simulated SNO data.  
In Section IV we describe our simulation of the SNO experiment and the 
subsequent data analysis. In Section V we critically examine if the zenith angle distribution 
at SNO adds to what can be learned from the D\&N spectra. 
We compare our expectations for SNO with an analysis of the zenith-angle 
distribution at Super-Kamiokande.
We summarize our results in \mbox{Section VI}.

\section{Day and Night Recoil Electron Energy Spectra}

The SNO CC data will provide an accurate determination of the shape of the energy spectrum 
 from  $^8$B neutrinos. 
Information about the oscillation parameters will be embedded in the overall suppression 
of the 
CC rate relative to that of the SSM and in the distortion of the shape of the energy 
spectrum. The 
low Q-value of the CC reaction \mbox{(1.442 MeV)} 
makes this process well-suited to obtaining a
spectrum with high energy resolution because most of the energy of the incoming neutrino is
carried away by the outgoing electron \mbox{($0 \leq T_e \leq E_{\nu}-Q$).} Since SNO is a real-time
experiment, it is capable of studying the effect of the Earth on neutrinos that pass through it
en route to the detector. A nadir angle, $\theta_Z$, is defined as the angle between the negative
$z$ axis of the coordinate system at the detector and the direction of the Sun. 
 With this definition, ${\rm{cos}}\,\theta_Z  \leq 0$ during the day and 
${\rm{cos}}\,\theta_Z  > 0$ at night. Conventionally, $\theta_Z$ is called the 
zenith angle although it is actually 
the complement of the zenith angle. The relative  amount of time the detector is exposed to the Sun at a particular
zenith angle is given by the zenith-angle exposure function~\cite{expose}.

Two electron energy spectra can be measured, one each for neutrinos detected in the daytime 
and nighttime. 
Each electron spectrum is divided into 19 bins, every 0.5 MeV from the kinetic energy threshold 
$T_{th}=5$ MeV{\footnote{Note that the target threshold for the CC reaction is 5 MeV kinetic 
energy, 
not total energy. We thank E. Beier for emphasizing this point.}} to 14 MeV and a last
bin that includes all events with energies from \mbox{14 MeV} to 20 MeV. 
The expectation in each day/night bin defined 
by $\Delta T_i \equiv [T_i^{min}, T_i^{max}]$ is
\begin{equation}
R_i^{D,N}={{\mathcal{N}}}\int_{0}^{\infty}dE_{\nu}\,\bigg(\Phi_{B}(E_{\nu}) +
1.8416 \times 10^{-3}\,\Phi_{hep}(E_{\nu}) \bigg)\,P^{D,N}(E_{\nu}) \,
\sigma_{CC}(E_{\nu},\Delta T_i)\,
\label{rate} 
\end{equation} 
where $\Phi_{B}$ and $\Phi_{hep}$ are the normalized energy spectra of the $^8$B and $hep$ 
neutrinos respectively. For the undistorted spectrum shape of the $^8$B neutrinos, 
we have adopted the 
result based on a measurement of the $\beta$-delayed $\alpha$ spectrum from the decay
of $^8$B~\cite{ortiz}, with this spectrum normalized to the flux of BPB2000. 
The factor $1.8416 \times 10^{-3}$ in~(\ref{rate}) is the relative total flux of 
$hep$ neutrinos to $^8$B neutrinos in the SSM (BPB2000). 
This factor is 4.5 times larger than that of 
BBP98~\cite{ossm} for two reasons: (i) A recent calculation of the $hep$ neutrino 
flux~\cite{marcucci} updates the BBP98 value by a factor of 4.4. (ii) In BPB2000,
the $^8$B flux is $5.05 \times {10^6}\, (1^{+0.20}_{-0.16})\,{\rm{cm}}^{-2}\,{\rm{s}}^{-1}$
versus $5.15 \times {10^6}\, (1^{+0.19}_{-0.14})\,{\rm{cm}}^{-2}\,{\rm{s}}^{-1}$ of BBP98.
%\begin{enumerate}
%\addtolength{\itemsep}{-2.5mm}
%\item{A recent calculation of the $hep$ neutrino 
%flux~\cite{marcucci} updates the BBP98 value by a factor of 4.4.}
%\item{In BPB2000,
%the $^8$B flux is $5.05 \times {10^6}\, (1^{+0.20}_{-0.16})\,{\rm{cm}}^{-2}\,{\rm{s}}^{-1}$
%versus $5.15 \times {10^6}\, (1^{+0.19}_{-0.14})\,{\rm{cm}}^{-2}\,{\rm{s}}^{-1}$ of BBP98.}
%\end{enumerate} 
The overall normalization ${{\mathcal{N}}}$ yields the expected
number of events in the absence of oscillations if the $\nu_e$ survival probability at 
the detector, $P^{D,N}(E_{\nu})$, is unity. If oscillations occur~\cite{mikh}, 
\begin{eqnarray}
P^{D}(E_{\nu})&=&  P_{\odot}(E_\nu)\,,\\
P^{N}(E_{\nu})&=&  P_{\odot}(E_\nu)+{1-2\,P_{\odot}(E_\nu) \over {\rm{cos}}2\,\theta}\,
(\vev{P_{e2}^{N}(E_\nu)} - {\rm{sin}}^2\theta)\,,
\end{eqnarray}
where $P_{\odot}$ is the probability that a neutrino
leaves the Sun as $\nu_e$, given by the well-known Parke formula~\cite{parke},
\begin{equation}
P_{\odot}(E_{\nu})={1\over 2}+\bigg({1\over 2}-P_c\bigg)\,{\rm{cos}}2\,\theta \,{\rm{cos}}2\,\theta^0_m\,,
\end{equation}  
where $\theta^0_m$ is the mixing angle in matter at the point of neutrino production in the
Sun. 
It is given by
\begin{equation}
{\rm{tan}}2\,\theta^0_m={{\rm{tan}}2\,\theta \over 1 - 
{2\sqrt{2}G_F N^0_e E_{\nu} \over \Delta m^2 {\rm{cos}}2\,\theta}}.
\end{equation}  
Here, $N^0_e$ is the electron density in the Sun at the creation point of the neutrino. 
An analytic expression for the crossing probability, $P_c$, 
which is a measure of the non-adiabaticity of the transitions is~\cite{pet},
\begin{equation}
P_c={\rm{exp}\big(-{\pi \over 2}\,\gamma\, F \big)-
\rm{exp}\big(-{\pi \over 2}\,\gamma\, {F \over {\rm{sin}}^2\,\theta} \big) \over 1-
\rm{exp}\big(-{\pi \over 2}\,\gamma\, {F \over {\rm{sin}}^2\,\theta} \big)}\,,
\end{equation}  
where $\gamma$ characterizes the adiabaticity of the resonance and
$F=1-{\rm{tan}}^2\theta$ for the exponentially varying matter density in the Sun. The
adiabaticity parameter is given by~\cite{parke,haxton}
\begin{equation}
\gamma={\Delta m^2 {\rm{sin}}^2 2\,\theta \over 2\,E_{\nu}\,{\rm{cos}}2\,\theta
|\dot{N_e}/N_e|_R}\,,
\end{equation}  
with $|\dot{N_e}/N_e|_R$ evaluated at the resonance.
 Finally, $\vev{P_{e2}^{N}}$ is the time-averaged
probability of the transition $\nu_2 \rightarrow \nu_e$ due to the effect of Earth 
matter{\footnote{We follow the usual conventions that $\nu_1$ and $\nu_2$ are the 
mass eigenstates with masses $m_1$ and $m_2$, the mass-squared difference,
$\Delta m^2\equiv m_2^2-m_1^2$, is positive and $\theta$ is the vacuum mixing angle which
can take values between 0 and $\pi/2$, thereby accommodating an inverted mass 
hierarchy~\cite{dark}.}}. We
assume the Preliminary Reference Earth Model~\cite{prem} for the Earth's electron density. 
The reduced cross section for producing an electron with measured kinetic energy in the 
interval $\Delta T_i$ is
\begin{equation}
\sigma_{CC}(E_{\nu},\Delta T_i) = \int_{T_i^{min}}^{T_i^{max}}dT \int_0^{T^{\prime}_{max}}dT^{\prime}\,
 {d\sigma_{CC} \over dT^{\prime}}(E_{\nu},T^{\prime}) R(T,T^{\prime}) \,,                 
\end{equation}  
where $d\sigma_{CC}/dT^{\prime}$ is the differential cross section for the
CC reaction (from~\cite{res}) with  $T^{\prime}$ being the actual kinetic energy of the electron. 
$T^{\prime}_{max}$ is the kinematic limit $E_{\nu}-Q$. $R(T,T^{\prime})$ 
is the energy resolution function that describes the distribution of the measured energy $T$
about the actual energy $T^{\prime}$ and is given by~\cite{res}
\begin{equation}
R(T,T^{\prime})={1 \over w\sqrt{2\,\pi\,T^{\prime}}}{\rm{exp}}
\bigg[- {(T-T^{\prime})^2 \over 2\,w^2\,\,T^{\prime}}\bigg]  
\label{reso}
\end{equation}
with $w=0.348$ MeV.

\section{Using the Similarities of Super-Kamiokande and SNO}

Super-Kamiokande and SNO are fairly similar experiments insofar as CC 
measurements are concerned. Both are high-statistics real-time 
electronic experiments
using Cerenkov light detection and both are sensitive to only $^8$B and 
$hep$ neutrinos because their energy thresholds are almost the same.
SK uses $H_2 O$ as its 
detection medium while SNO uses $D_2 O$. Thus, SK detects
solar neutrinos via the elastic scattering (ES) reaction
\begin{equation}
\nu_x + e^- \rightarrow \nu_x' + e^-\ \ \ \ \ \ (\rm{ES})\,,
\end{equation}
which only makes a minor contribution to the rate at SNO. However, since the neutrino source 
and the principle of neutrino detection 
are the same in 
both experiments, it is reasonable to expect the two experiments to yield 
equivalent flux measurements. This equivalence has been exploited to devise ways to predict the
NC rate at SK once SNO has CC rate results~\cite{villante} and 
to predict the energy spectrum at SNO from that measured by SK~\cite{equal}.

We assume that the equivalence of SK and SNO is sufficiently robust 
that the
best $\chi^2$ definition for SK will also be the best for SNO.
 The calculation for the ES rate at SK is similar to
the CC rate described above for SNO except for the following alterations:
\begin{enumerate}
\addtolength{\itemsep}{-2.5mm}
\item{$P\,\sigma_{CC}$ is replced by  $P\,\sigma_{e}+(1-P)\,\sigma_{\mu}$, where
$\sigma_{e}$ and $\sigma_{\mu}$ are the $\nu_{e}-e$ and $\nu_{\mu}-e$
ES cross sections~\cite{kamion}, respectively.}
\item{ The bins are defined in terms of the total electron 
energy, since SK reports its data in terms of the reconstructed total energy of the 
recoil electron, with threshold $E_{\nu}=5$ MeV.}
\item{In Eq.~(\ref{reso}), $w=0.47$ MeV~\cite{sk}. }
\end{enumerate} 

 SK has reported results from 1258 days of data-taking~\cite{sk,s} 
as ratios with respect
to the first version of BPB2000 in which the $^8$B flux is 
 \mbox{$5.15 \times {10^6}\, (1^{+0.20}_{-0.16})\,{\rm{cm}}^{-2}\,
{\rm{s}}^{-1}$}~\cite{ssm2}. 
We call this
$\rm{SSM}^{\prime}$. A recently revised version of BPB2000 gives the $^8$B flux as
\mbox{$5.05 \times {10^6}\, (1^{+0.20}_{-0.16})\,{\rm{cm}}^{-2}\,{\rm{s}}^{-1}$}~\cite{SSM}. 
Consequently, we modify
the SK data accordingly, by multiplying the central value and statistical error 
in each bin by the ratio $5.15/5.05$. The systematic errors are conveniently given as percentages and
do not need modification. The measured flux suppression is~\cite{sk,s}
\begin{equation}
{{\rm{Data_{SK}}}\over{\rm{SSM}^{\prime}}}=0.451^{+0.017}_{-0.015}\,,
\end{equation} 
which relative to the SSM is
\begin{equation}
{{\rm{Data_{SK}}}\over{\rm{SSM}}}=0.459^{+0.018}_{-0.016}\,.
\label{ratio}
\end{equation} 
SK has presented results using several different $\chi^2$ definitions.
In their latest flux-independent analysis of the D\&N spectra, they used~\cite{s}
\begin{equation}
{\chi^2_{SK}(\Delta m^2,{\rm{tan}^2}\theta)} = \sum_{i=1}^{38}\bigg[
{(\phi_i^{meas}/\phi_i^{SSM'}-\alpha\,f_i(\beta)\,\phi_i^{osc}
/\phi_i^{SSM'})^2 \over (\sigma_i^{stat})^2+(\sigma_i^{uncorr})^2}
+\bigg({\beta \over \sigma_i^{corr}}\bigg)^2\,\bigg],
\label{chisk}
\end{equation} 
where the flux measured by SK in
the $i^{\rm{th}}$ bin 
is $\phi_i^{meas}$, the expected flux without oscillations is 
$\phi_i^{SSM'}$ and the expected flux with oscillations is 
$\phi_i^{osc}\equiv \phi_i^{osc}(\Delta m^2,{\rm{tan}^2}\theta)$. The uncertainties
$\sigma_i^{stat}$, $\sigma_i^{uncorr}$ and $\sigma_i^{corr}$ are the 
statistical, uncorrelated and correlated uncertainties in the $i^{\rm{th}}$ 
bin, respectively. The correlated errors include the experimental uncertainties
in the determination of the $^8$B spectrum~\cite{ortiz} and the theoretical
uncertainties in the calculation of th expected energy spectrum~\cite{err}. 
The functions, $f_i(\beta)$, 
parameterize the correlated uncertainty in the shape of
the spectra; $\beta$ is a free shift factor of the correlated error and
 $\alpha$ is a free parameter that normalizes the measured flux relative to the 
expected flux. The sum runs over 38 energy bins (19 day bins + 19 night bins). 
 
\begin{figure}[ht]
\mbox{\psfig{file=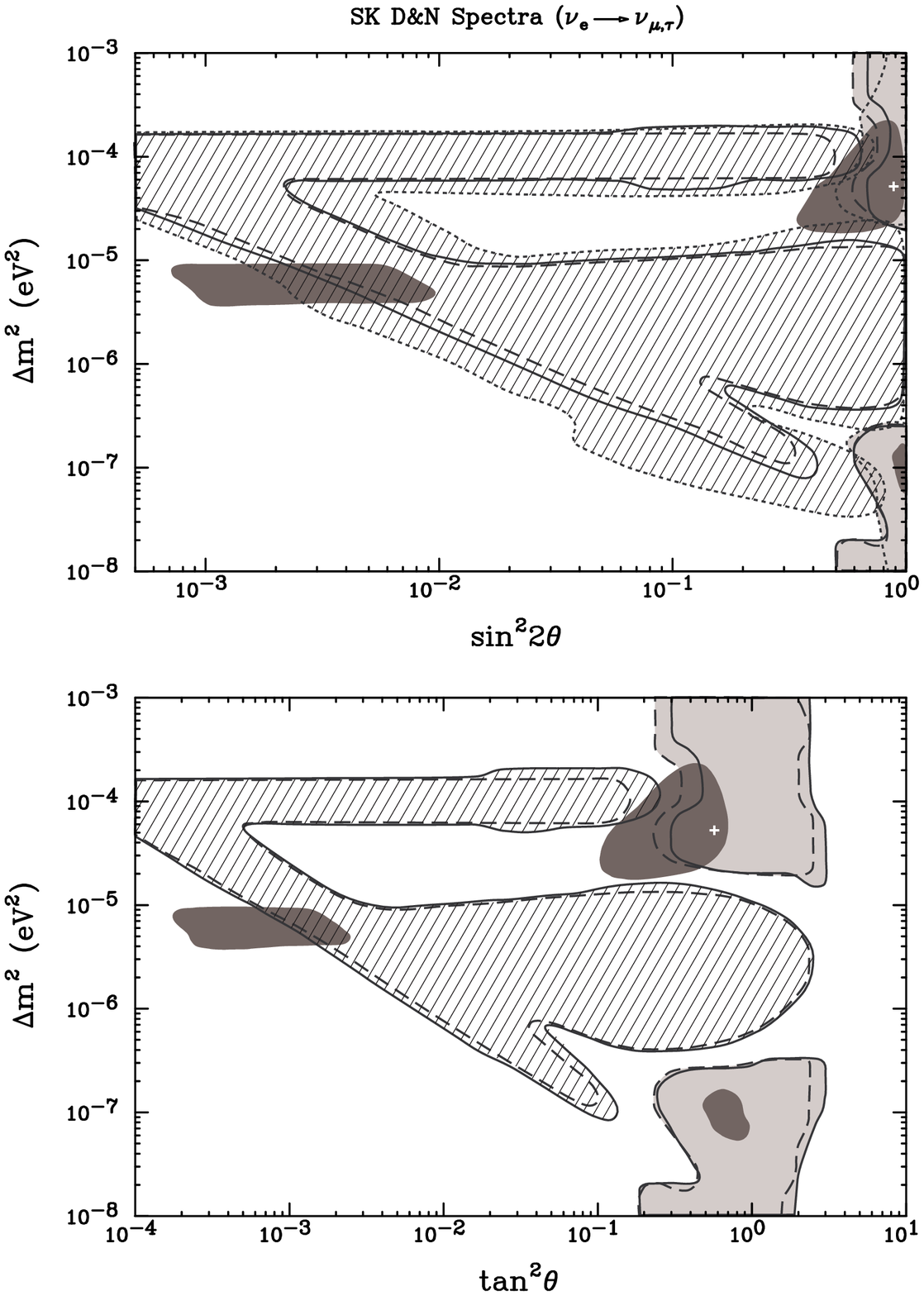,width=15cm,height=17cm}}
\caption[]{The exclusion (hatched) and allowed (lightly shaded) regions at 95\% C. L. 
obtained from the 1258 day SK D\&N energy spectra using three 
different $\chi^2$ definitions. 
The regions enclosed by the dotted (taken from Ref.~\cite{s}), 
dashed and solid lines result from the 
use of Eqs.~(\ref{chisk}),(\ref{con}) and (\ref{statchi}), respectively. 
Equation~(\ref{statchi}) has no contribution from systematic errors. The flux constraint 
is imposed to find the allowed regions. The crosshairs mark the best-fit parameters, 
$(\Delta m^2,{\rm{tan}^2}\theta)=(5.01\times 10^{-5}\,{\rm{eV}^2},0.60)$ using Eqs.
(\ref{con}) and (\ref{statchi}). 
The dark shaded regions are the global 
solutions (with 1117 SK days) at \mbox{99\% C. L.} with free $^8$B and $hep$ fluxes 
found in Ref.~\cite{analysis}. }
\label{skspex}
\end{figure}

We have taken the SK 95\% C. L. exclusion region from Ref.~\cite{s} for our later 
comparison with regions obtainable with alternative $\chi^2$ definitions.
It is the hatched region enclosed by the dotted line in Fig.~\ref{skspex}. 
Note that this exclusion region corresponds to data
relative to $\rm{SSM}^{\prime}$. The dark shaded areas are the allowed regions at 99\% C. L. 
from a global analysis with free $^8$B and $hep$ fluxes and the SSM 
(not $\rm{SSM}^{\prime}$)~\cite{analysis}. The LOW solution is allowed only at the 
99\% C. L.. The analysis 
includes the D\&N spectra from the 1117-day event sample 
but not the total rate, since that information
 is contained in the energy spectra. Ref.~\cite{analysis} used the shape
of the undistorted spectrum of $^8$B neutrinos from Ref.~\cite{err}.  
We emphasize that although the SMA region found from a combined fit
of flux measurements is excluded at the 95\% C. L.~\cite{s}, the SMA region from the 
global fit of Ref.~\cite{analysis} is not.

Another suitable definition of $\chi^2$, similar to one used by SK in earlier analyses 
(of 825 effective days of data), is
\begin{equation}
{\chi^2(\Delta m^2,{\rm{tan}^2}\theta)} = \sum_{i=1}^{38}\bigg[
{\big(\phi_i^{meas}/\phi_i^{SSM}-\alpha/(1+\beta\,\sigma_i^{corr})\,\phi_i^{osc}
/\phi_i^{SSM}\big)^2 \over (\sigma_i^{stat})^2+(\sigma_i^{uncorr})^2}\bigg]
+\beta^2\,,
\label{con}
\end{equation}
 Again, $\alpha$   
is a free flux normalization factor and $\beta$ constrains
the variation of correlated systematic errors. 
Performing a $\chi^2$ analysis of the 
1258 day data with this 
definition gives the 95\% C. L. exclusion (hatched) region ($\chi^2>52.19$ for 37 d.o.f.), 
outlined by the dashed line in Fig.~\ref{skspex}. 
The $hep$ contribution to the
neutrino flux is left unconstrained in the SK analysis while we have fixed the ratio 
between the  $hep$ and $^8$B fluxes. Thus, for a test of the MSW hypothesis, the 
SK analysis has 36 degrees of freedom while our analysis has 37.  
Also, keeping in mind that the regions enclosed by the dotted and dashed lines correspond
to two different reference solar models and $\chi^2$ definitions, it
is noteworthy that the general shapes of the regions closely
resemble each other, although they differ in size as a consequence of the term 
$\sum\, (\beta/\sigma_i^{corr})^2$ in $\chi^2_{SK}$. 
Even if $\beta$ is very small, $\beta \sim 0.001$, the contribution from
this term can significantly increase the value of $\chi^2_{SK}$ thereby permitting a larger
exclusion region\footnote{If we set $f_i(\beta)\equiv 1$, thereby making the flux
normalization the same in all bins, and $\beta=0.001$, we improve the agreement with 
the SK region significantly.}.  On the other hand, by including the correlated errors in 
as in~(\ref{con}), their effect is greatly diminished.
It is evident that the spectral distortion
functions $f_i(\beta)$ play an important role in defining an efficient $\chi^2$ function. 
To include the possibility
of negative $\Delta m^2$, 
we have also plotted the same region (dashed line) 
with ${\rm{tan}^2}\theta$ as the abscissa~\cite{dark}.

In the approximation that systematic errors can be neglected,
both the above $\chi^2$ definitions lead to  
\begin{equation}
{\chi^2_{stat}(\Delta m^2,{\rm{tan}^2}\theta)} = \sum_{i=1}^{38}\bigg[
{\phi_i^{meas}/\phi_i^{SSM}-\alpha\,\phi_i^{osc}
/\phi_i^{SSM} \over \sigma_i^{stat}}\bigg]^2
\,.
\label{statchi}
\end{equation}
The resulting 95\% C. L. exclusion region (hatched and enclosed by the solid line) 
is shown in Fig.~\ref{skspex}. 
Again, we note the remarkable similarity of the shapes of the exclusion regions from the three
analyses. Dropping systematic
errors leads to a region more similar in size to that obtained by the SK collaboration than
that obtained by using Eq.~(\ref{con}).  
On this basis, we hereafter assume that we can safely
ignore all systematic errors when making projections for SNO with simulated data. 
 If SK is a reasonable guide, we will err on the conservative side. 
It may be counterintuitive
that the exclusion region using $\chi^2_{SK}$ is larger than that using 
$\chi^2_{stat}$ because
one expects more errors to lead to less confidence and therefore a smaller exclusion region.
However, as explained earlier, the correlated errors in Eq.~\ref{chisk} are responsible 
for this.
The region using Eq.~(\ref{con}) is smaller than that of $\chi^2_{stat}$, in
agreement with expectations.

So far we have only considered flux-independent exclusion plots. If instead, 
flux-dependent allowed regions are sought, one needs to add another term to the 
$\chi^2$ definitions considered,
\begin{equation}
{\chi^2(\Delta m^2,{\rm{tan}^2}\theta)}\longrightarrow {\chi^2(\Delta m^2,{\rm{tan}^2}\theta)}+
\bigg({1-\alpha \over \sigma_{\alpha}}\bigg)^2\,,
\label{stat}
\end{equation}
where $\sigma_{\alpha}=^{+0.20}_{-0.16}$SSM (or $\rm{SSM}^{\prime}$) 
is the theoretical uncertainty in the $^8$B flux. In our analysis we symmetrize this
value to $\sigma_{\alpha}=\pm 0.18\,$SSM. The 95\% C. L. allowed regions 
($\Delta \chi^2 <5.99$ for two oscillation parameters), are 
superimposed on the exclusion plots in Fig.~\ref{skspex}. 
The allowed region from 
SK's analysis is not shown in Ref.~\cite{s}; we have taken it from Ref.~\cite{smy}.
 It is evident that the
allowed regions are alike with minor differences in size. 
 With the $\chi^2$-definitions of Eqs.~(\ref{con}) and (\ref{statchi}), we  
find the same best-fit parameters, 
$(\Delta m^2,{\rm{tan}^2}\theta)=(5.01\times 10^{-5}\,{\rm{eV}^2},0.60)$ with 
 $\chi^2=30.6$ using Eq.~(\ref{con}) and $\chi^2=32.4$ using Eq.~(\ref{statchi}) for 
36 degrees of freedom. 

In Fig.~\ref{skspex}, the crosshairs represent the best-fit 
parameters which are very close to those presented by SK from an analysis of the data 
from 1117 days in which they include the
flux constraint~\cite{2000}; SK does not report the best-fit point from a
flux-dependent analysis of the 1258-day D\&N spectra. 
From a flux-independent analysis they find
that the minimum $\chi^2$ value lies in the VAC region~\cite{s}, 
which we have not considered in our analysis. However, from a flux-dependent 
analysis of their zenith spectrum, they find the minimum $\chi^2$ in the 
VAC region, with some points in the LMA region with similar $\chi^2$ values~\cite{s}. 
For example, 
$(\Delta m^2,{\rm{tan}^2}\theta)=(7 \times 10^{-5}\,{\rm{eV}^2},0.47)$ is one such point, 
which is close to our best-fit parameters.
A cautionary note when interpreting 
Fig.~\ref{skspex} is that the dark-shaded flux-independent globally allowed regions of 
Ref.~\cite{analysis} are not directly comparable to our flux-dependent allowed regions. Our
motivation for superimposing the flux-independent allowed regions is to facilitate a 
comparison with the flux-independent exclusion regions. 

\section{Data Simulation and Analysis}

If the SSM flux is correct, then in the absence of oscillations SNO should detect about
9250 events per year. If instead the SSM flux is wrong and no oscillations occur, 
the flux suppression expected at SNO is the same as that seen by
SK, Eq.~(\ref{ratio}). 
We first simulate data assuming this pessimistic scenario and predict the exclusion and 
allowed regions that we can expect SNO to present three years from now. 
Then we turn to the more reasonable explanation in which neutrino oscillations do occur. 

We 
simulate data for the best-fit oscillation parameters in each of the three allowed MSW 
solutions (from the global analysis of Ref.~\cite{analysis}) and display
the corresponding exclusion and allowed regions. Exclusion regions are meaningful 
only when the data points are normally distributed and that for at least
some region of the parameter space of interest, $\chi^2$/d.o.f $\sim 1$. We enforce this
 by simulating data for which this is true at the input oscillation parameters.
The normalization constant ${{\mathcal{N}}}$ in Eq.~(\ref{rate}) is set by the stipulation that
we are considering three years of data accumulation. 

\begin{figure}[ht]
\mbox{\psfig{file=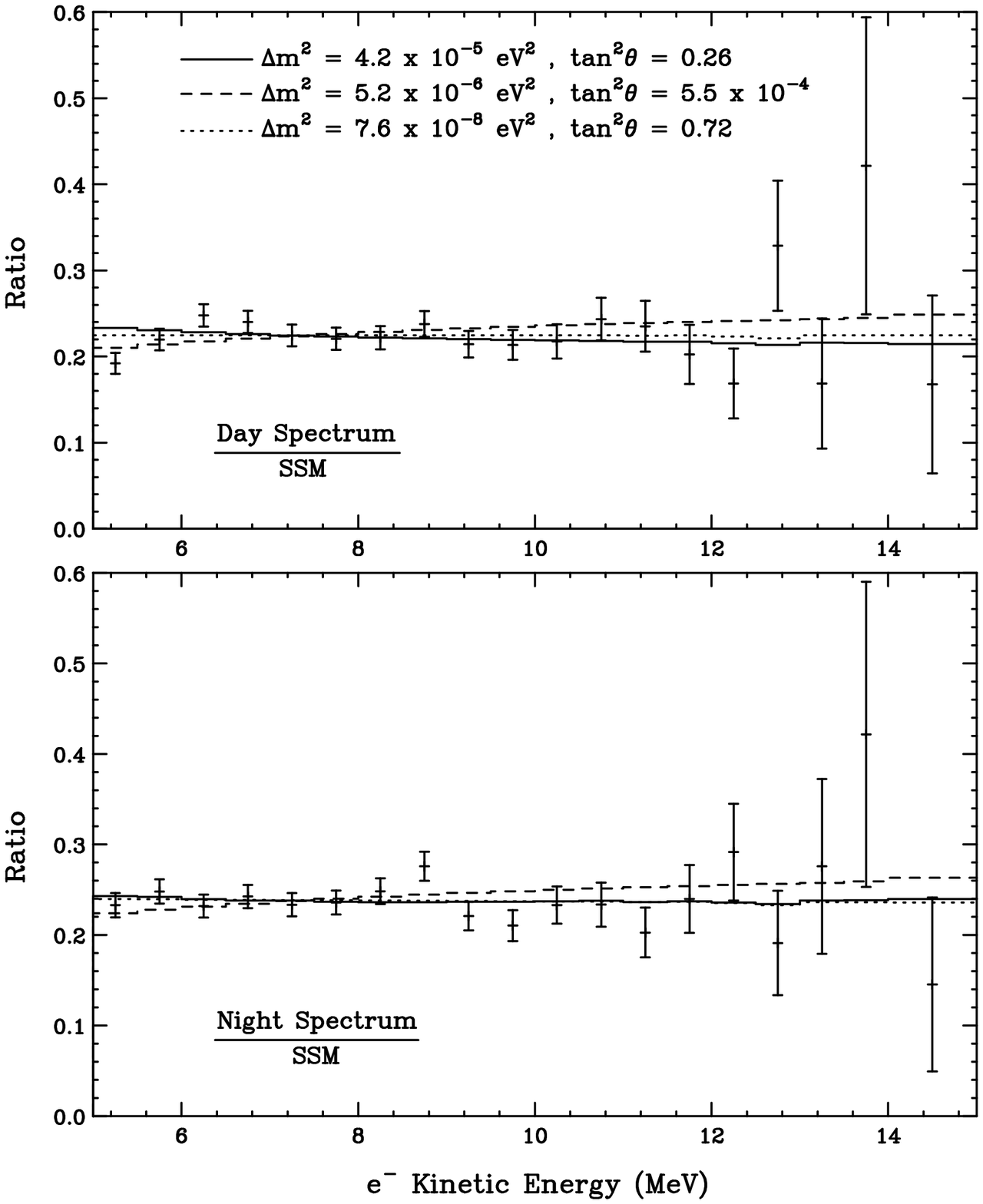,width=15cm,height=17cm}}
\caption{The expected $e^-$ D\&N energy spectra at 
SNO (as a ratio with respect to the SSM) 
with three years of accumulated data for typical LMA (solid histogram), 
SMA (dashed histogram) and LOW (dotted histogram) solutions. 
The data points are simulated for the LMA solution. The spectra are normalized
to the flux reflected by the data. The last
bin includes all energies from \mbox{14 MeV} to 20 MeV. }
\label{snospec}
\end{figure}

Figure~\ref{snospec} shows the expected 
spectra (as a ratio with respect to the SSM) for typical LMA (solid histogram), 
SMA (dashed histogram) and LOW (dotted histogram) solutions. 
The data points here are simulated for the LMA solution. For direct comparison of the 
spectral shapes, these spectra are normalized
to the flux corresponding to the simulated data.

For the sake of specificity, we define $\chi^2$ as
\begin{equation}
{\chi^2(\Delta m^2,{\rm{tan}^2}\theta)} = \sum_{i=1}^{38} 
\bigg[{R^{simulated}_i/R^{SSM}_i-\alpha\, 
R_i^{osc}/R^{SSM}_i \over \sigma_i^{stat}}\bigg]^2 \,,
\label{s}
\end{equation}
where $\sigma_i^{stat}=\sqrt{R^{simulated}_i}/R^{SSM}_i$ and  
the number of simulated events, $R^{simulated}_i$, in bin $\Delta T_i$ is obtained by randomly choosing a point 
from a Gaussian 
distribution centered at the theoretical
value and of width equal to the square root of the theoretical value. This definition is 
the same as that of Eq.~(\ref{stat}) except that it is expressed 
in terms of the number of events rather than the flux.

\begin{figure}[ht]
\mbox{\psfig{file=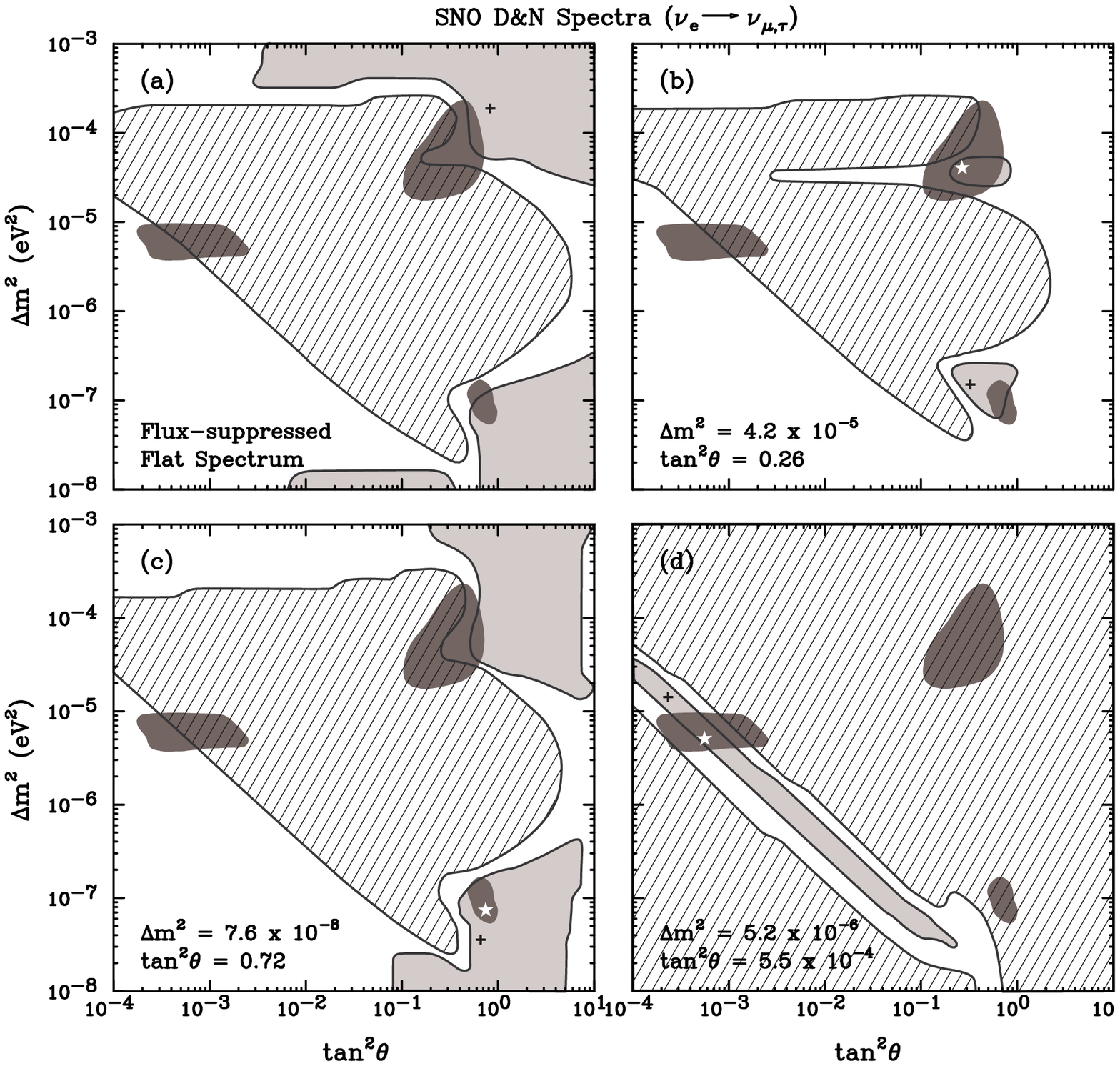,width=16.5cm,height=17cm}}
\caption[]{Expected 99\% C. L. exclusion (hatched) and allowed (lightly shaded) regions from
the D\&N spectra at SNO for 
(a) a flux-suppressed flat spectrum (b) LMA solution, (c) LOW solution and 
(d) SMA solution with three years of accumulated data. The dark shaded regions are the global 
solutions at 99\% C. L. with free $^8$B and $hep$ fluxes. The stars and crosshairs mark the
theoretical inputs and best-fit points, respectively. }
\label{snoregions}
\end{figure}

To be conservative, we only show 99\% C. L. exclusion ($\chi^2>59.89$ for 37 d.o.f.), 
and allowed regions ($\Delta \chi^2 <9.21$),
 resulting from the simulated SNO data. Figure~\ref{snoregions} 
shows the expected regions for the 
flat spectrum hypothesis (with the same flux suppression as seen by SK) and 
three MSW solutions. The stars and crosshairs mark the
theoretical inputs and best-fit points, respectively.
The SMA expectation is characteristic and easily identifiable.
For all other possibilities, the plots bear a striking semblance to each other and 
to the SK results. The LMA and LOW solutions persist simultaneously for all but the SMA 
solution. The 
SMA region is excluded to a large extent and with a more efficient
$\chi^2$ definition, the entire SMA region could be excluded.
%In Fig...., we present the expectation 

\section{Zenith Angle Distribution}

In principle the information contained in the zenith angle distribution is contained
in the D\&N spectra. The SK exclusion regions obtained by them in separate 
analyses of the D\&N spectra 
and the zenith
angle distribution (with energy subdivisions) bear testimony to this expectation; 
the differences
are small~\cite{s}. This is despite the fact that each zenith 
angle bin is split into several energy bins, thereby potentially maximizing the resolution 
available to SK. It has been advocated that an appropriate choice of binning might make it
 possible for SNO to not only identify which solution is correct but also to determine the
oscillation parameters~\cite{snozenith}. By performing a complete analysis, we now assess the
extent to which this claim can be validated.

The choice of
night bins of equal size in cos$\,\theta_Z$ is democratic, but does not take advantage of any 
distinctive features of the distributions of the different solutions. 
In the context of SNO, the qualitative behavior of the distributions for the LMA, SMA and LOW 
solutions was studied in detail in Ref.~\cite{snozenith}.
 It was found that with a suitable choice of binning,
 any smearing of peculiarities intrinsic to the region of parameter space can be avoided. 
The events were binned in a manner that leads to a characteristic distribution for the 
LOW solution because in the SMA and LMA regions, the cos$\,\theta_Z$-dependence is rather weak leading to a more or less flat distribution. 
This remark is  pertinent, since as we have seen,
it is difficult to differentiate the LMA and LOW solutions from each other. 
If the SMA solution is the correct one, the strong spectral distortion will easily make it 
stand apart. 
As defined in Table~\ref{bins}, there is one day bin and five non-uniform night bins.
The ``core bin'', N5, at SNO ((cos$\,\theta_Z)_{max}=0.92$) is smaller than that at SK 
((cos$\,\theta_Z)_{max}=0.975$) because SNO's higher latitude restricts the $\theta_Z$ 
range. Thus, a
smaller number of solar 
neutrinos that pass through the Earth's core are incident at SNO than at SK. 
\begin{table}[t]
\begin{eqnarray}
\begin{array}{lrcccl}
\rm{Data set}      &   & & & &  \\
\hline
\rm{Day}           &     -1 &\leq &  {\rm{cos}}\,\theta_Z  & \leq &  0               \\  
\rm{N1}            &      0 & <  & {\rm{cos}}\,\theta_Z  & \leq &  0.173               \\
\rm{N2}            &     0.173 & < & {\rm{cos}}\,\theta_Z &  \leq &  0.5                   \\
\rm{N3}            &     0.5 & < &   {\rm{cos}}\,\theta_Z & \leq & 0.707                  \\
\rm{N4}            &     0.707 & < & {\rm{cos}}\,\theta_Z & \leq &  0.83                   \\
\rm{N5}            &     0.83 & < &  {\rm{cos}}\,\theta_Z & \leq &  0.92                  \nonumber
\end{array}
\end{eqnarray}
\caption[]{The definitions of the night bins in terms of the nadir of the Sun, $\theta_Z$. Note 
that the ``core bin'', N5, does not contain zeniths beyond 0.92 because the latitude of the
detector restricts its range.}
\label{bins}
\end{table}

Since we know the zenith-angle
exposure function~\cite{expose}, the SSM prediction for the number of events 
  in each zenith angle bin, $R_{i,SSM}^Z$, and the prediction with oscillations, 
$R_{i,\,osc}^Z$, can be calculated.
 For a given set of oscillation parameters, we want to generate zenith angle 
distributions that have the same number of events 
as the simulated energy spectra of the previous section. 
The number of events in the day bin is simply 
the sum of all the events in the day spectrum. For the night bins, we use 
$R_{i,\,osc}^Z$ as the central value of a Gaussian distribution and simulate the number
of events in each night bin. Note that
 the number of events at night represented by this distribution does not 
coincide with that of the night spectrum. The nighttime distribution is renormalized
to yield the number of events in the night spectrum. 
Now the simulated energy spectra and zenith angle distribution
reflect the same data. To match the number of simulated events in the zenith angle
distribution and the D\&N spectra, two normalizations are needed, one each for the
daytime and nighttime events. The shape of the theoretical expectation, 
however involves only one
normalization, namely, the total number of simulated events.

% The top panel of Fig.~\ref{snozen} shows the expected zenith angle distributions 
%with three years of accumulated data for typical LMA (solid histogram), 
%SMA (dashed histogram) and LOW (dotted histogram) solutions. 
%The simulated LMA data reflects the data for the D\&N spectra in 
%Fig.~\ref{snospec}.
%To facilitate easy comparison of the shapes of the distributions, they are normalized to 
%the total simulated flux. 
%$2.32 \times {10^6}\,{\rm{cm}}^{-2}\,{\rm{s}}^{-1}$, corresponding to a flux suppression
%of 0.459.  It may be unwise to perform a {\it chi-by-eye} analysis, 
%but it is unlikely that either the LMA or LOW solution can be isolated. 
%\begin{figure}[ht]
%\mbox{\psfig{file=/u/bwood/sno/snozen.ps,width=15cm,height=17cm}}
%\caption{The top panel shows the expected zenith angle distributions at SNO for the 
%three sets of oscillation parameters used for the D\&N spectra analysis. 
%The data points correspond to the same dataset as in Fig.~\ref{snospec}. 
%The distributions are normalized
%to the flux corresponding to the dataset. 
%The bottom panel shows 99\% C. L. exclusion regions for 
%datasets simulated for each of the MSW solutions in the top panel.}
%\label{snozen}
%\end{figure}
\begin{figure}[ht]
\mbox{\psfig{file=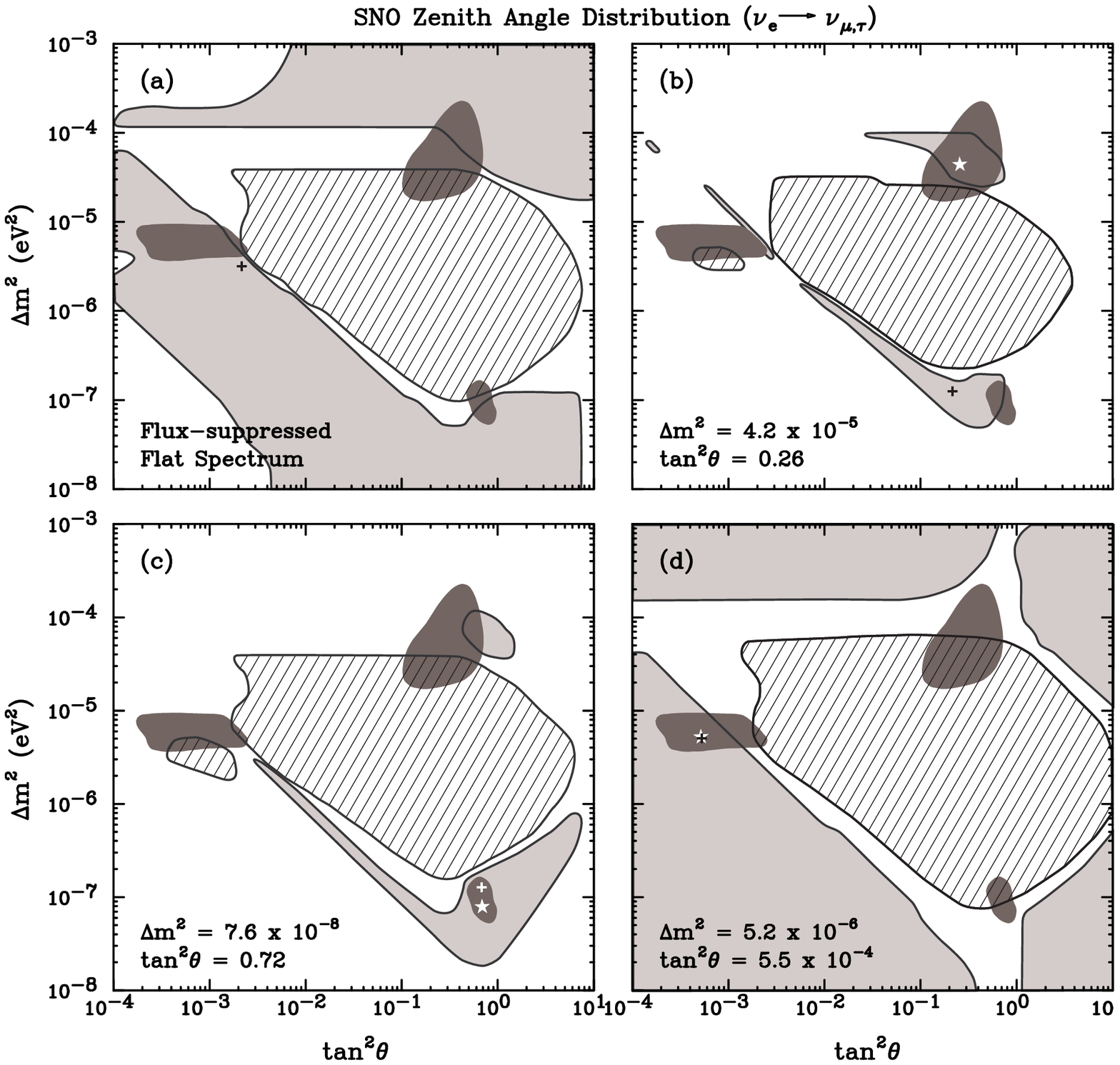,width=16.5cm,height=17cm}}
\caption[]{Expected 99\% C. L. exclusion (hatched) and allowed (lightly shaded) regions from 
zenith angle distributions at SNO corresponding to the simulated datasets of 
Fig.~\ref{snoregions}.  The stars and crosshairs mark the
theoretical inputs and best-fit points, respectively. 
  }
\label{snozex}
\end{figure}

We perform flux-independent analyses 
(for the same datasets used to find the regions of Fig.~\ref{snoregions}) 
with the simple $\chi^2$ function,
\begin{equation}
{\chi^2(\Delta m^2,{\rm{tan}^2} \theta)}^Z = \sum_{i=1}^{6}\, 
\bigg[{R_{i,\,simulated}^Z/R_{i,\,SSM}^Z-\alpha\,
R_{i,\,osc}^Z/R_{i,\,SSM}^Z \over \sigma_i^Z}\bigg]^2 \,,
\label{z}
\end{equation}
where $\sigma_i^Z=\sqrt{{R_{i,simulated}^Z}}/R_{i,SSM}^Z$. From 
Fig~\ref{snozex}, it is evident that the zenith angle 
distributions of the various solutions lead to similar 99\% C. L. exclusion regions 
($\chi^2>15.09$ for 5 d.o.f.). 
We underscore the fact that the regions corresponding to the flat spectrum, LMA and 
LOW datasets exclude part of the SMA region, and that the region found with the SMA dataset
excludes a large part of the LMA and LOW solutions. This is consistent with 
Fig.~\ref{snoregions}. Additionally, 
the 99\% C. L allowed regions have the same shapes for the
LMA and LOW solutions.  The flux constraint is included to find  
the allowed regions. The stars and crosshairs mark the
theoretical inputs and best-fit points, respectively. 

\begin{figure}[ht]
\mbox{\psfig{file=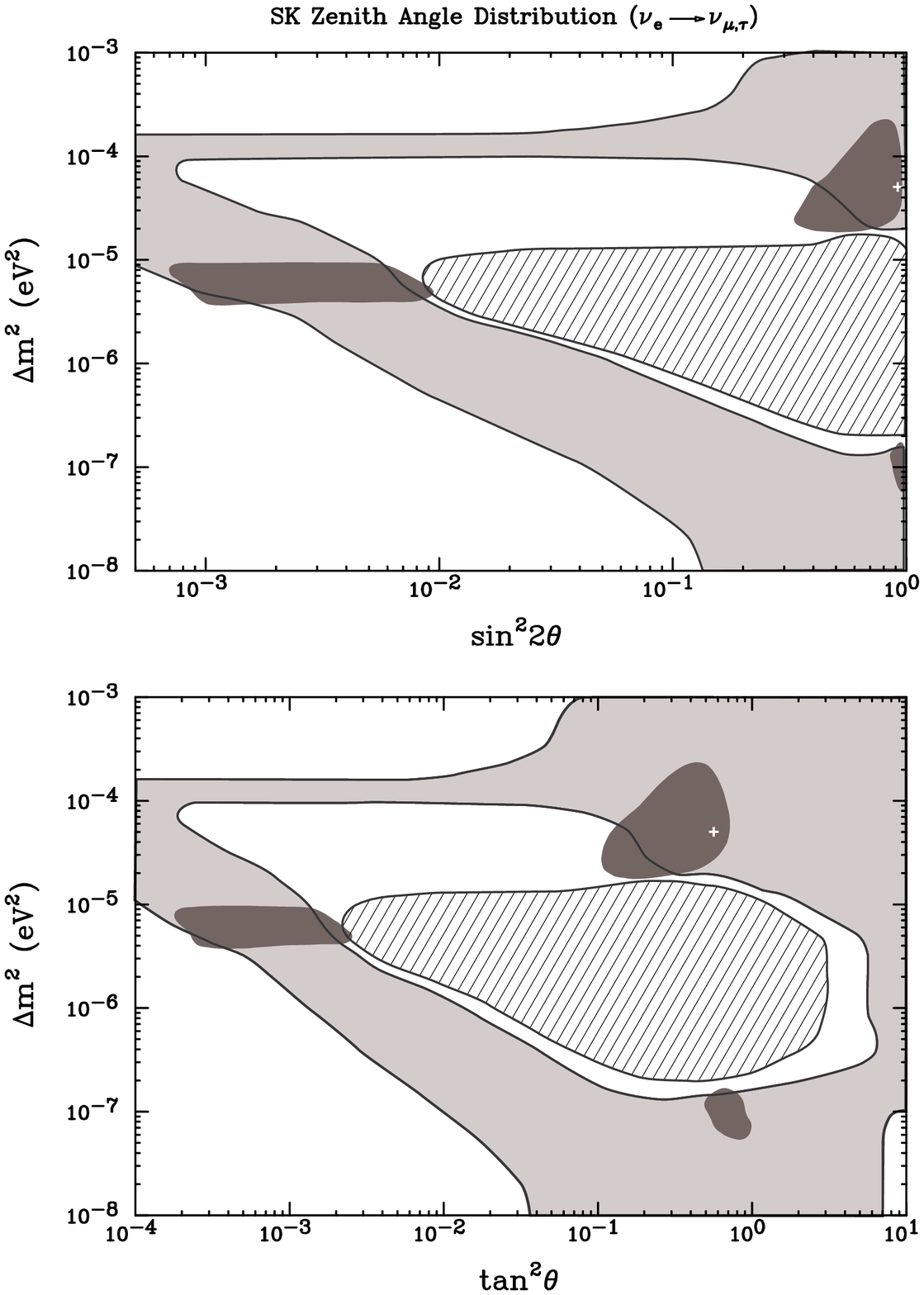,width=15cm,height=17cm}}
\caption{The exclusion (hatched) and allowed (lightly shaded) regions at 99\% C. L. 
from the 1258-day SK zenith angle distribution. The crosshairs mark the best-fit parameters, 
$(\Delta m^2,{\rm{tan}^2}\theta)=(5.01\times 10^{-5}\,{\rm{eV}^2},0.56)$.    }
\label{skzen}
\end{figure}

Figure~\ref{skzen} shows the 99\% C. L. exclusion ($\chi^2>16.81$ for 6 d.o.f.),
and allowed regions from the
zenith angle distribution from 1258 days of data at SK. In making these regions, we
 have employed SK's binning  
which is different from our choice for SNO. 
 The statistical
and systematic errors are added in quadrature. The crosshairs mark the best-fit parameters,  
$(\Delta m^2,{\rm{tan}^2}\theta)=(5.01\times 10^{-5}\,{\rm{eV}^2},0.56)$ with $\chi^2=5.15$.
The exclusion regions of \mbox{Figs.~\ref{snozex}--\ref{skzen}} 
are similar to the 99\% C. L. exclusion region reported by SK (with a 504-day dataset)
with a similar $\chi^2$ definition~\cite{skz}. No parts of the globally allowed regions
are excluded.  
Of course, subdividing each zenith angle bin into energy bins (as done by SK in their
latest analysis~\cite{s}) will
greatly improve the sensitivity of the analysis, but this is equivalent to using 
 D\&N energy spectra.

The day-night variation embodied in the zenith angle distribution can also be presented 
in terms of a day-night asymmetry defined by
\begin{equation}
A_{DN}=2\, {N-D \over N+D}=2\, {P^N-P^D \over P^N+P^D}\,,
\end{equation}
where $D$ and $N$ are the total number of events detected during the days and nights, respectively. The approximate ranges of $A_{DN}$ are $(0.005,0.1)$ in the LMA and LOW regions and  
$(-0.01,0.05)$ in the SMA region.
Note that in parts of the SMA region, $A_{DN}<0$, thus uniquely identifying the SMA solution. 
However, $A_{DN}>-0.01$
 and an identification of such a small deviation from zero will be difficult.
Semi-analytic approximations for $A_{DN}$ have been derived for the three allowed 
regions~\cite{anal}. These 
expressions can be used to provide insight into the orientations of the zenith-angle 
allowed regions.
Iso-asymmetry lines in $A_{DN}$ that pass through all three regions must have $A_{DN}>0$ and are given by
\begin{eqnarray}
\Delta m^2\, ({\rm{eV}^2})&\approx& {3 \times 10^{-6} }\, {{\rm{sin}}^2 2\,\theta \over  A_{DN}}
\ \ \ \ \ \ \ \ (\rm{LMA})\,, \label{dnlma}\\
A_{DN} &\approx&  {\rm{sin}}^2 2\,\theta\ \ \ \ \ \ \ \ \ \ \ \ \ \ \, \ \ \ \ \ \ (\rm{SMA})\,, 
\label{dnsma}\\
\Delta m^2\, ({\rm{eV}^2})&\approx& 2.5 \times 10^{-6}\,
{A_{DN} \over {\rm{sin}}^2 2\,\theta}\ \ 
\ \ \ \ (\rm{LOW})\,.
\label{dnlow}
\end{eqnarray}
 The relations in Eqs.~(\ref{dnlma}--\ref{dnlow}) have a wider domain of applicability than
indicated. For example, Eq.~(\ref{dnlow}) is applicable in the range $10^{-8} \lsim 
\Delta m^2 \lsim 3 \times 10^{-6}$. This explains why the 99\% C. L. allowed regions almost
connect the SMA and LOW solutions. 

\section{Summary}

We have analyzed the 1258-day day and night energy spectra presented by Super-Kamiokande 
using $\chi^2$ definitions that account for systematic errors in different ways. The best-fit
lies in the LMA region at 
$(\Delta m^2,{\rm{tan}^2}\theta)=(5.01\times 10^{-5}\,{\rm{eV}^2},0.60)$, independently of
whether systematic errors are included in the $\chi^2$-definition. We have
shown that these approaches lead 
to exclusion and allowed regions of different sizes, but the
general areas of the regions remain unchanged even if systematic errors are
neglected (see Fig.~\ref{skspex}). 
Using Super-Kamiokande as our reference, we then
draw conclusions for SNO based
on analyses that incorporates only statistical errors. 
We assume the optimistic electron kinetic energy threshold of 5 MeV and three years of 
accumulated data.  

If the SMA solution is correct, the day and night 
spectra will show sufficiently strong distortions to distinguish it from 
the other solutions (Fig.~\ref{snoregions}(d)). 
For a flux-suppressed flat spectrum or the LMA and LOW 
solutions, the regions are similar enough to not provide any constraint beyond the
exclusion of most of the the SMA solution (Fig.~\ref{snoregions}(a--c));
 the LMA and LOW solutions are indistinguishable. However, 
KamLAND~\cite{kamland} will certainly help in this regard by either ruling out the LMA
solution or pinning down the LMA oscillation parameters~\cite{kam}.
The zenith angle distribution in itself will not add 
anything to what can be obtained from the day and night spectra 
unless each zenith angle bin is
subdivided into energy bins. 

The 99\% \mbox{C. L.} 
exclusion regions in Fig.~\ref{snozex}  bear a striking similarity
to that obtained from the 1258-day zenith angle distribution at 
Super-Kamiokande (Fig.~\ref{skzen}), for which the best-fit parameters  are $(\Delta m^2,{\rm{tan}^2}\theta)=(5.01\times 10^{-5}\,{\rm{eV}^2},0.56)$.
 We expect results from charged-current measurements at SNO to be similar to that of
Super-Kamiokande, thus providing an important check of the Super-Kamiokande conclusions.
Needless to say, neutral current data from SNO will provide a crucial test of the 
existence of sterile neutrinos, and
if solar neutrinos do not oscillate to sterile neutrinos, SNO will measure
the $^8$B flux produced in the Sun.

\vspace{0.25in}
\newpage
\acknowledgements 
We thank E. Beier, M. Gonzalez-Garcia, P. Krastev, J. Learned, Y. Takeuchi, 
M. Smy and Y. Suzuki for useful inputs.
This work was supported in part by the U.S.~Department of Energy
under Grants No.~DE-FG02-94ER40817 and No.~DE-FG02-95ER40896,
and in part by the Wisconsin Alumni Research Foundation.
\vspace{0.25in}

%\newpage

\end{document}